# DISCRETE SELF-SIMILARITY BETWEEN RR LYRAE STARS AND SINGLY-EXCITED HELIUM ATOMS

Robert L. Oldershaw

Earth Sciences Building

Amherst College

Amherst, MA 01002

rloldershaw@amherst.edu

**Running Head**: RR LYRAE STARS AND HELIUM ATOMS

**Abstract**: Classical variable stars called RR Lyrae stars have pulsating outer envelopes constituted of excited atoms. Here we demonstrate that the qualitative and quantitative properties of RR Lyrae variables and one subclass of their atomic scale constituents: singly-excited helium atoms undergoing transitions between Rydberg states, share a remarkable degree of self-similarity. In terms of masses, radii, oscillation periods, morphologies and kinematics the stellar and atomic analogues obey a simple set of discrete self-similar scaling equations. The concept of stellar/atomic self-similarity may prove useful in the search for a deeper understanding of both stellar and atomic systems.

1. INTRODUCTION

Fractal structures are common to virtually all realms of nature.[1] Galaxy distributions, fluid turbulence, topographic shapes, neuronal interconnections and the clustering of stars or plasma particles represent a small sampling of the broad domains wherein fractal phenomena are ubiquitous. A key property of fractal systems is their self-similarity, in which similar morphological or temporal patterns recur on different size or time scales throughout the hierarchical structure of the system. Here we demonstrate a surprisingly robust self-similarity between the classical pulsating stars called RR Lyrae variables and one subclass of their atomic scale constituents: helium atoms undergoing transitions between adjacent energy levels.

2. PRELIMINARIES

At the outset of this investigation we define a heuristic set of discrete self-similar scaling equations that allow us to correlate observed stellar mass, radius and oscillation period values with experimental measurements for helium atoms in Rydberg states. Initially we will treat these scaling equations as axioms, deferring a discussion of their origin until the end of our investigation, except to say that they were derived decades ago without reference to RR Lyrae stars. The scaling equations are:

$$R \approx \Lambda r \qquad (1)$$

$$P \approx \Lambda p \qquad (2)$$

$$M \approx \Lambda^D m \,. \qquad (3)$$

R, P and M are radii, periods and masses of RR Lyrae stars; r, p and m are the counterpart parameters of helium atoms; and $\Lambda$ and **D** are dimensionless scaling constants equal to $5.2 \times 10^{17}$ and 3.174, respectively.

RR Lyrae variables[2,3] are blue giant stars with classifications of A or F. They are thought to pulsate primarily in the fundamental radial mode (l = 0), and $(R_{max} - R_{min}) \div R_{min} \approx 10\%$. It has been "definitely established" that the oscillation takes place in the outer envelope of the star, rather than in its core.[2] RR Lyrae stars tend to oscillate with a single period, but cases of double-mode pulsation with $P_1/P_0 \approx 0.746$ are not uncommon. Although the periods of RR Lyrae variables range from $\approx 0.2$ days to $\approx 1.0$ days, the overwhelming majority have periods between 0.25 days and 0.75 days. Three subtypes have been identified: RRc variables with nearly sinusoidal light curves and periods of roughly 0.3 days, RRa variables with asymmetric light curves and periods of roughly 0.5 days and RRb variables (often combined with the RRa class) with intermediate light curve asymmetry and periods of roughly 0.7 days.[3] The most typical mass for RR Lyrae stars is $<M> \approx 0.6\ M_\odot$ and their radii range from approximately 3.7 $R_\odot$ to 7.2 $R_\odot$.[4]

Given the *approximate* value of $<M>$, we find that $<M> \approx \Lambda^D m_{He}$, in agreement with Eq.3 to within a factor of 0.045. Given the stellar radius range of 3.7 $R_\odot$ to 7.2 $R_\odot$ and Eq. 1, we can estimate that the self-similar radius range for the $^4$He atom should be $4.95 \times 10^{-7}$ cm to $9.64 \times 10^{-7}$ cm. Then using the general radius versus principal quantum number (n) relation[5], $r \approx 2n^2 a_o$, for Rydberg atoms with low angular momentum quantum numbers (l), where $a_o$ is the Bohr radius of $\approx 0.53 \times 10^{-8}$ cm, we can determine that the relevant range of n values is roughly 6.8 to 9.5, which rounds off to $7 \leq n \leq 10$. Since RR Lyrae stars appear to be mostly radial mode oscillators, we assume that the relevant range of l values for $^4$He is $0 \leq l \leq 1$. We also make the tentative assumption that the most likely atomic scale transitions for $^4$He in 1sns or 1snp states with $7 \leq n \leq 10$ are single-level transitions, i.e., $\Delta n = 1$.

3. QUANTITATIVE TEST

With the above preliminaries completed, we are now ready to put the putative example of discrete cosmological self-similarity to a crucial test: RR Lyrae oscillation periods should match up uniquely with the specified $^4$He transition periods when the latter are scaled in accordance with Eq.2. Table 1 presents the $^4$He data[6] needed for this test, and the predicted RR Lyrae periods derived from that data.

**TABLE 1** Transition Data for $^4$He (1sns and 1snp), Singlet and Triplet States, $7 \leq n \leq 10$, $\Delta n = 1$, and Predicted RR Lyrae Periods

| $n_1 \rightarrow n_2$; $^x$S | $\Delta E$ (atomic units) | Transition Period $1/\nu$ (sec) | Predicted RR Lyrae Oscillation Period (days) |
|---|---|---|---|
| 8p $\rightarrow$ 7s; $^3$S | 0.00318 | 4.7725 x 10$^{-14}$ | **0.2872** |
| 8p $\rightarrow$ 7s; $^1$S | 0.00284 | 5.3559 x 10$^{-14}$ | **0.3223** |
| 8s $\rightarrow$ 7s; $^3$S | 0.00270 | 5.6275 x 10$^{-14}$ | **0.3387** |
| 8s $\rightarrow$ 7s; $^1$S | 0.00253 | 6.0056 x 10$^{-14}$ | **0.3614** |
| | | | |
| 9p $\rightarrow$ 8s; $^3$S | 0.00216 | 7.0340 x 10$^{-14}$ | **0.4233** |
| 9p $\rightarrow$ 8s; $^1$S | 0.00194 | 7.8446 x 10$^{-14}$ | **0.4722** |
| 9s $\rightarrow$ 8s; $^3$S | 0.00183 | 8.3029 x 10$^{-14}$ | **0.4997** |
| 9s $\rightarrow$ 8s; $^1$S | 0.00172 | 8.8339 x 10$^{-14}$ | **0.5317** |
| | | | |
| 10p $\rightarrow$ 9s; $^3$S | 0.00153 | 9.9130 x 10$^{-14}$ | **0.5966** |
| 10p $\rightarrow$ 9s; $^1$S | 0.00138 | 1.0991 x 10$^{-13}$ | **0.6615** |
| 10s $\rightarrow$ 9s; $^3$S | 0.00129 | 1.1778 x 10$^{-13}$ | **0.7089** |
| 10s $\rightarrow$ 9s; $^1$S | 0.00123 | 1.2353 x 10$^{-13}$ | **0.7435** |

If the concept of discrete self-similarity between RR Lyrae stars and $^4$He atoms has merit, then the 12 predicted oscillation periods listed in Table 1 should be identifiable in RR Lyrae period distributions. As an initial indication of the general agreement between predictions and observations, we note that the 12 periods can be subdivided naturally into 3 subgroups representing the n = 8→7, n = 9→8 and n = 10→9 transitions. The average oscillation periods for these three subgroups is roughly 0.3 days, 0.5 days and 0.7 days, which corresponds rather well with the observed average periods for RRc, RRa and RRb stars, respectively. Also, note that there are potentially diagnostic "gaps" in the distribution of predicted periods at roughly 0.37 – 0.41 days, 0.43 – 0.46 days, 0.54 – 0.58 days and 0.61 – 0.65 days. To achieve a more rigorous quantitative test, we need a sizeable sample of RR Lyrae variables that is reasonably homogeneous and analyzed with care. Fortunately an excellent test sample has recently become available.

One of the secondary benefits of microlensing searches for stellar-mass dark matter objects has been the discovery of substantial numbers of variable stars. The Optical Gravitational Lensing Experiment (OGLE) team has recently published a catalog[7] of RR Lyrae stars found within the Large Magellanic Cloud galaxy. The period data were subjected to narrow binning and multiple-binning analyses, yielding period distributions that surpass the accuracy of previous period histograms. Figure 1 shows the distribution of periods for 84 RR Lyrae variables found in the rich star cluster NGC 1835.

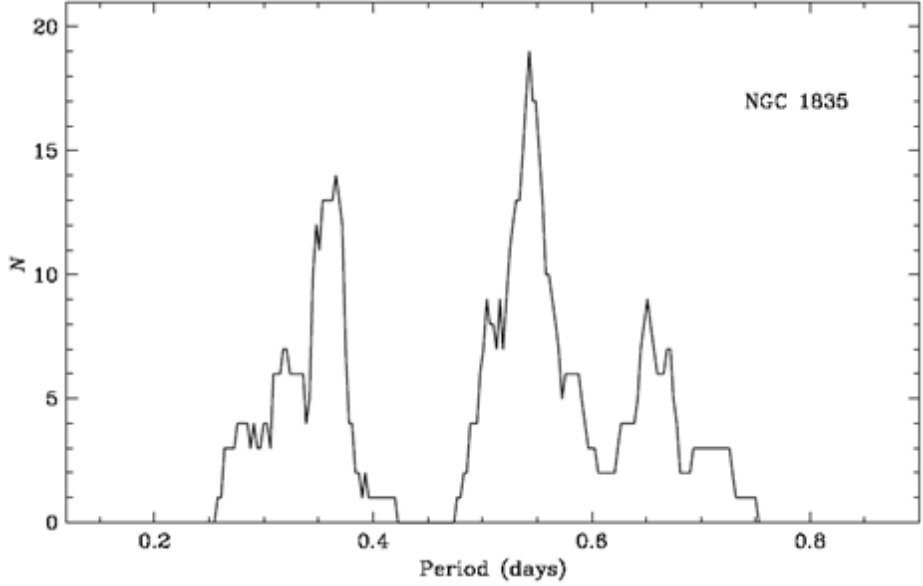

**FIGURE 1**  Period Distribution for RR Lyrae Stars in NGC 1835[7].

One can identify 10 structural features in this distribution: 6 peaks, 2 "shoulders", 1 gap and 1 "valley". Each of these structural features matches up, to within 1% or better, with one of the periods or gaps that were uniquely predicted solely on the basis of the physical properties of $^4$He (1sns,p) transitions with $7 \leq n \leq 10$ and $\Delta n = 1$. In Table 2 we summarize these results.

**TABLE 2**  Observed and Predicted Periods (NGC 1835)

| Observed Feature (days) | $^4$He Transition | Predicted Period (days) | Error (days) |
|---|---|---|---|
| **0.28** [peak] | 1s8p – 1s7s ($^3$S) | **0.2872** | 0.007 |
| **0.32** [peak] | 1s8p – 1s7s ($^1$S) | **0.3223** | 0.002 |
| **0.36** [peak] | 1s8s – 1s7s ($^1$S) | **0.3614** | 0.001 |
| **0.50** [peak/shoulder] | 1s9s – 1s8s ($^3$S) | **0.4997** | 0.0003 |
| **0.54** [peak] | 1s9s – 1s8s ($^1$S) | **0.5317** | 0.008 |
| **0.59** [shoulder] | 1s10p – 1s9s ($^3$S) | **0.5966** | 0.007 |
| **0.66** [peak] | 1s10p – 1s9s ($^1$S) | **0.6615** | 0.002 |
| **0.71** [peak] | 1s10s – 1s9s ($^3$S) | **0.7089** | 0.001 |
| **0.44** [gap] |  | **0.43 – 0.46** | 0 |
| **0.62** [valley] |  | **0.61 – 0.65** | 0 |

4. CAVEATS

It is important to acknowledge the physical factors that tend to shift atomic transition frequencies away from their unperturbed values, or add spurious frequencies to the test spectra. In the case of $^4$He (1sns,p) transitions with $7 \leq n \leq 10$ and $\Delta n = 1$, there is the expected doubling of lines due to a splitting that permits singlet or triplet systems. Sets of spurious periods will be introduced into the test spectra if systems with $l > 1$ (and therefore $-l \leq m \leq +l$) "contaminate" the test sample. Likewise, if systems other than helium but in He-like configurations such as $Li^+$ (1sns,p) or $H^-$ (1sns,p) are present, then additional sets of spurious transition periods will be detected. If $^3$He and $^6$He isotopes are present in the test sample, then still more sets of spurious transition periods will be observed. Lastly with regard to spurious periods, systems undergoing transitions with $\Delta n > 1$ can add further sets of transition periods to the spectra.

Turning now to shifts away from the unperturbed positions of transition frequencies or periods, there are at least four physical factors that can cause significant shifts. Due to the relatively large charge separations of Rydberg configurations, ambient *electric fields* can cause substantial energy level shifts. Likewise, the energy levels of Rydberg atoms are susceptible to significant shifting due to their sensitivity to ambient *magnetic fields*. *Temperature* and *pressure*, which are often quite high in astrophysical settings, can also broaden and shift peaks in the test spectra.

Given these four sources of spurious periods and the four physical causes of line shifting, which would tend to be present at some level in *non-laboratory settings*, we should maintain reasonable expectations for the discreteness of non-laboratory period distributions of $^4$He atoms and the self-similar period distributions of RR Lyrae stars.

Moreover, given the stochastic nature of quantum phenomena, we would expect different samples of either type of systems to show significant variability in the numbers of specific transition periods present.

5. CONCLUSIONS

Starting from a knowledge of the physical properties of RR Lyrae variable stars and a set of tentative self-similar scaling equations, we have demonstrated that RR Lyrae stars and $^4$He atoms (1sns,p; $7 \leq n \leq 10$) undergoing energy level transitions with $\Delta n = 1$ are quantitatively self-similar in terms of masses, radii and oscillation periods. The stellar and atomic analogues share the qualitative properties of being spherical systems undergoing energetic pulsations of limited duration that take place in their outer envelopes. An interesting question for future research is whether this remarkable self-similarity can be extended to include other classes of variable stars such as ZZ Ceti, W Virginis, $\delta$ Scuti, Cepheid and Mira variables.

Finally, it should be mentioned that the self-similar scale transformation equations used in this paper were developed in 1985 for a discrete fractal paradigm called the Self-Similar Cosmological Paradigm (SSCP).[8-11] The SSCP emphasizes nature's intrinsic and well-stratified hierarchical organization, proposing that the hierarchy is divided into discrete cosmological Scales, of which we can currently observe the Atomic, Stellar and Galactic Scales. The SSCP also proposes that the discrete Scales are rigorously self-similar to one another, such that for each class of fundamental particle, composite system or physical phenomenon on any given Scale there are self-similar analogues on all other Scales. At present the number of Scales cannot be known, but for reasons of natural philosophy it is tentatively proposed that there are a denumerably infinite number of Scales, ordered in terms of their intrinsic ranges of space, time and mass scales. The spatial (R), temporal (T) and mass (M) parameters of discrete self-similar analogues on neighboring Scales $\Psi$ and $\Psi$-1 are related by the following set of discrete self-similar Scale transformation equations.

$$R_\Psi = \Lambda R_{\Psi-1} \quad , \quad (4)$$

$$T_\Psi = \Lambda T_{\Psi-1} \quad \text{and} \quad (5)$$

$$M_\Psi = \Lambda^D M_{\Psi-1} \quad , \quad (6)$$

where $\Lambda$ and D are empirically determined dimensionless scale factors equal to $5.2 \times 10^{17}$ and 3.174, respectively. The value of $\Lambda^D$ is $1.70 \times 10^{56}$. The symbol $\Psi$ is used as an index to distinguish different Scales, such that

$$\Psi = \{\ldots, -2, -1, 0, +1, +2, \ldots\} \quad , \quad (7)$$

and the Stellar Scale is usually assigned $\Psi = 0$. Thus, Atomic Scale systems and phenomena are designated by $\Psi = -1$ and Galactic Scale systems and phenomena are assigned $\Psi = +1$. The fundamental self-similarity of the SSCP and the recursive character of the discrete scaling equations suggest that nature is an infinite discrete fractal, in terms of its morphology, kinematics and dynamics.  Perhaps the most thorough and accessible resource for exploring the SSCP is the author's website.[12]